\documentclass[conference]{IEEEtran}
\IEEEoverridecommandlockouts
\usepackage[utf8]{inputenc}
\usepackage{subfigure,times,graphics,mathptm,epsfig,amsmath,xspace,endnotes,pifont,multirow,rotating,listings,amssymb,algorithmic,color,caption,nicefrac,adjustbox,todonotes,tabularx,mathtools, algorithmic,algorithm}
\usepackage{graphicx}
\newtheorem{definition}{Definition}

\usepackage{tabularx}
\newcolumntype{L}[1]{>{\raggedright\arraybackslash}p{#1}} 
\newcolumntype{C}[1]{>{\centering\arraybackslash}p{#1}} 
\newcolumntype{R}[1]{>{\raggedleft\arraybackslash}p{#1}} %

\usepackage{amsmath}
\usepackage{amssymb}
\usepackage{graphicx}
\usepackage{nccmath}
\usepackage{accents}
\usepackage{stackengine}
\usepackage{tikz}
\usepackage{textcomp}

\author{Nariman Torkzaban, and John S. Baras\\
\tt\small Department of Electrical and Computer Engineering\\
and the Institute for Systems Research\\
University of Maryland, College Park, MD 20742, USA\\
Email: \tt\small \{ narimant | baras \} @umd.edu \\
}
\newcommand\copyrighttext{%
  \footnotesize \textcopyright 2020 IEEE. Personal use of this material is permitted.
  Permission from IEEE must be obtained for all other uses, in any current or future
  media, including reprinting/republishing this material for advertising or promotional
  purposes, creating new collective works, for resale or redistribution to servers or
  lists, or reuse of any copyrighted component of this work in other works.
  }
\newcommand\copyrightnotice{%
\begin{tikzpicture}[remember picture,overlay]
\node[anchor=south,yshift=10pt] at (current page.south) {\fbox{\parbox{\dimexpr\textwidth-\fboxsep-\fboxrule\relax}{\copyrighttext}}};
\end{tikzpicture}%
}
\begin{document}
%
\title{Trust-Aware Service Function Chain Embedding: A Path-Based Approach}



\maketitle
\copyrightnotice

\begin{abstract}
With the emergence of network function virtualization (NFV), and software-defined networking (SDN), the realization and implementation of service function chains (SFCs) have become much easier. An SFC is an ordered set of interconnected virtual network functions (VNFs). NFV allows for decoupling the network functions from proprietary hardware realizing a software-based implementation of VNFs on commodity hardware, and SDN decouples the network control from its forwarding logic allowing for a more flexible and programmable traffic routing among the VNFs. 
The SFC embedding problem (i.e. placement of SFCs on a shared substrate and establishing the corresponding traffic routes between the VNFs), has been extensively studied in the literature. 

In this paper, we extend a previous work on trust-aware service chain embedding with generalizing the role of trust by incorporating the trustworthiness of the service network links and substrate network paths into the SFC embedding decision process. We first introduce and formulate the path-based trust-aware service chain embedding problem as a mixed integer-linear program (MILP), and then provide an approximate model based on selecting  $k-$ shortest candidate substrate paths for hosting each virtual link, to reduce the complexity of the model. We validate the performance of our methods through simulations and conduct a discussion on evaluating the methods and some operation trade-offs.




\end{abstract}


%
\IEEEpeerreviewmaketitle

\section{Introduction}

The recent advances in NFV and SDN has enabled the network operators to launch and manage their networks faster, easier, and cheaper. Accordingly, given the advanced virtualization and programmability features, there are less operations and consequently less cost associated with service presentation and maintenance. More specifically, NFV reduces the network provisioning cost by decoupling the network functions from the proprietary hardware and implementing them on commodity hardware. In contrast to traditional networks where various network functions such as firewall, deep packet inspection, intrusion detection systems, video optimizer, etc. where deployed using specialized hardware, instances of such VNFs can be implemented on virtual machines(VMs) or containers, allowing for an easier and a more flexible provisioning of scalable solutions and driving higher profitability for the network providers. 

Complex network service requests can be provisioned by service function chaining. An SFC is comprised from  an  ordered  set  of  inter-connected  virtual  network  functions  (VNFs) with logical dependencies. In this context, the SFC embedding problem (i.e.  placement  of  SFCs  on  a  shared NFV infrastructure  and  establishing the  corresponding  traffic  routes  between  the  VNFs) is of a great significance.

Inspired by the notion of trust in NFV \cite{incorp} and with the goal of integrating the NFV security requirements in SFC embedding decisions, the trust-aware SFC embedding problem was motivated and introduced in \cite{sds19}, by integrating trust weights in the SFC embedding problem, where the security demands of each NF and the trustworthiness level of each substrate host were represented by the trust weights. In this paper, we generalize the approach in \cite{sds19} by augmenting the role of trust. More precisely, we take into account the trustworthiness requirements for both the NFs and the edges between them. In a similar fashion we assign trust values to the substrate network paths as well as the substrate hosts to model the trustworthiness of the NFV infrastructure. Similar to \cite{sds19}, we assume such trust values are computed and aggregated by a trust evaluator process, based on the network configuration and monitoring data and are distributed and provided to the entity in charge of the SFC embedding (network orchestrator, or controller) in a timely manner. Interested readers are referred to \cite{baras6}, and \cite{sds19} for more detailed information on trust and its integration to the SFC embedding problem.

We also note that the method in \cite{sds19} is \textit{link-based}; i.e.
i) The flow decision variables are link-to-link; and ii) Provided in the output is the assignment of each request link to a set of substrate links that are guaranteed to generate  valid continuous substrate paths by suitable flow formation and conservation constraints.
However, in this paper we represent the SFC embedding problem by a \textit{path-based} model; i.e.  i) The flow decision variables are link-to-path; and ii) In the output, the request links are directly assigned to the pre-selected substrate paths. 

In fact, one of the main contributions of this paper is to propose a path-based model for the SFC embedding problem (PB-SCE) which provides multiple advantages over the traditional link-based formulation used in \cite{sds19}.  Firstly, a path-based formulation allows for the integration of various network and routing policies within the service chain embedding framework with low complexity. For instance, PB-SCE can be simply augmented by a path pre-selection  phase to admit requirements such as, traffic splitting, guaranteeing maximum delay or cost, or even assuring the existence of (disjoint) backup paths.  

Moreover, within the path-based framework many of the design metrics that would enforce non-linear constraints to the link-based formulation (e.g. reliability, trust, availability, etc.), can simply be computed along the network paths in an offline fashion and be input to the path-based formulation. For instance, it is not possible in the link-based formulation in \cite{sds19} to incorporate a linear constraint for capturing the trust requirement of each virtual edge; however, in the path-based model it is straightforward to compute the trustworthiness of a network path following the corresponding trust aggregation policy and then input it to the model as a linear constraint. This is the main motivation for introducing the path-based trust-aware SFC embedding (PB-TASCE) model.

Finally, we note that in the context of trust-aware service chain embedding, a pth-based model allows for abstracting out the method by which trustworthiness of the infrastructure is computed and aggregated. Precisely, considering a path-based approach which only requires the trust values assigned to the paths, disregarding how this value is computed based on the similar for the underlying components, allows for the application of our method in different settings, where trust needs to be modeled differently \cite{baras6}. For instance, interpreting trust as a multiplicative metric will lead to a different trustworthiness judgment for a path comparing to the case where trustworthiness of a path is computed as the minimum of the trustworthiness of all its edges.

The remainder of the paper is organized as follows. Section~\ref{sec:desc} describes the trust-aware service chain embedding problem. In Section~\ref{sec:problem} we introduce the path-based MILP formulation and its approximate $k-$shortest path based variants. Section \ref{sec:evaluation} presents our evaluation results, whereas Section \ref{sec:relatedwork} provides an overview of the related work. Finally, in Section \ref{sec:conclusions}, we highlight our conclusions and discuss directions for future work.

\section{Network Model and Problem Description} 
\label{sec:desc}

The substrate network, is modeled as an undirected graph $G_s = (N_s, E_s)$, while the request network is modeled as a directed graph  $G_f = (N_f, E_f)$. Each substrate node $u \in N_s$ has a residual processing capacity $r_u$, and each substrate link $(u,v) \in E_s$ has a bandwidth capacity of $c_{uv}$, while the CPU requirement of request node $i \in N_f$ and the bandwidth demand of a request link $(i,j) \in E_f$ are represented by $g^i$ and $d^{ij}$ accordingly.

We denote by $t_u$ the trustworthiness of the substrate node $u \in N_s$, and by $t^i$ the trust requirement of the request node $i \in N_f$, while this metric for a request link $(i,j) \in E_f$ is denoted by $t^{ij}$ and for a substrate path $p$ by $t_p$, where $p$ is a connected set of edges in the substrate graph. As in \cite{sds19}, trust takes fractional numerical value in $[0,1]$. We note that the trustworthiness of a substrate path can be any function (according to specific use-case or methodology) of the trust values corresponding to the links and nodes belonging to that path.  

We define the following components of the path-based formulation to facilitate the description of the model:
\vspace{2mm}

\begin{definition}
{\textbf{Augmented Graph. }}
For a commodity (virtual link) $k = (i,j)$ where $ij \in E_f$ we denote by $G^k_s = (N^k_s, E^k_s)$, the augmented graph corresponding to commodity $k$, whereby for every node $u \in N_s$ that is eligible for hosting request node $i$, the directed \textbf{augmented edge} $(i, u)$ is added to $E_s$. Similarly, for every node $u$ that is eligible for hosting the request node $j$, the directed \textbf{augmented edge} $(u,j)$ will be added to $E_s$. Hence, for the augmented graph, explicitly we will have:

$$ N^k_s = N_s  \cup \{i, j\} \quad$$ $$ E^k_s = E_s \cup \{iu | u\in N_s \quad and \quad t_u \geq t^i \quad and \quad r_u \geq g^i \}  $$ $$ \quad \quad \quad \quad \cup\{uj | u\in N_s \quad and \quad t_u \geq t^j \quad and \quad r_u \geq g^j \}$$

Furthermore, we denote by $G^a_s= (N^a_s, E^a_s)$ the augmented graph corresponding to the request graph $G_f = (N_f, E_f)$, which contains all the nodes and links in all of the augmented graphs for all the commodities. 
\label{def1}
\end{definition}

\vspace{2mm}

\begin{definition}
{\textbf{Augmented Path.}}
For a commodity $K = (i,j)$ where $(i,j) \in E_f$ we denote by $p^{k}$ from $i$ to $j$, a generic augmented path corresponding to commodity $k$, where the initial and the final links are augmented edges corresponding to commodity $k$. In other words, once we remove the initial and final edge from $p^k$ the result will be a path of the original graph $G_s$. We further denote by $\mathcal{P}^k$, the set of all augmented paths corresponding to commodity $k$, and by $\mathcal{P}$ the set of all augmented paths. 
\label{def2}
\end{definition}

\begin{figure*}[t]
\begin{center}
\begin{minipage}[h]{0.225\textwidth}
\includegraphics[width=1\linewidth]{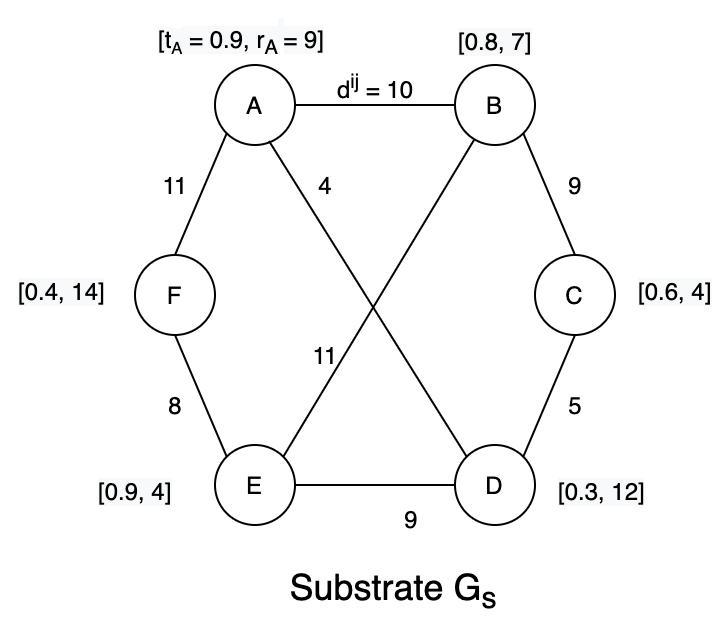}
\caption{ Substrate Graph $G_s = (N_s , E_s)$}
\label{fig:ssssub}
\end{minipage}
\hspace{1em}
\begin{minipage}[h]{0.225\textwidth}
\includegraphics[width=1\linewidth]{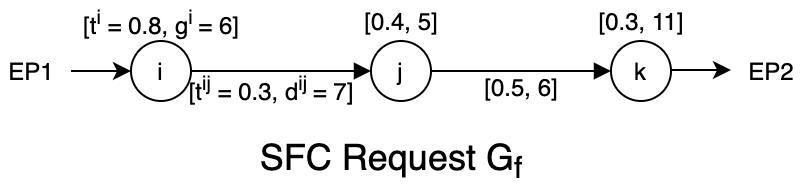}
\caption{Request Graph $G_f = (N_f , E_f)$ }
\label{fig:reeeeq}
\end{minipage}
\hspace{1em}
\begin{minipage}[h]{0.225\textwidth}
\includegraphics[width=1\linewidth]{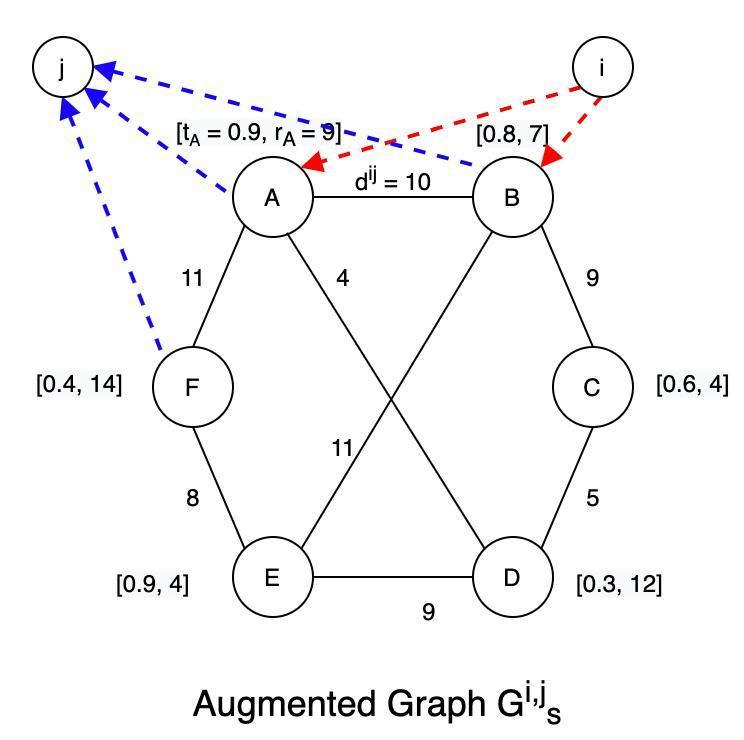}
\caption{ Augmented Substrate Graph $G^{ij}_s = (N^{ij}_s , E^{ij}_s)$ for virtual link $ij$ }
\label{fig: reqaug}
\end{minipage}
\hspace{1em}
\begin{minipage}[h]{0.225\textwidth}
\includegraphics[width=1\linewidth]{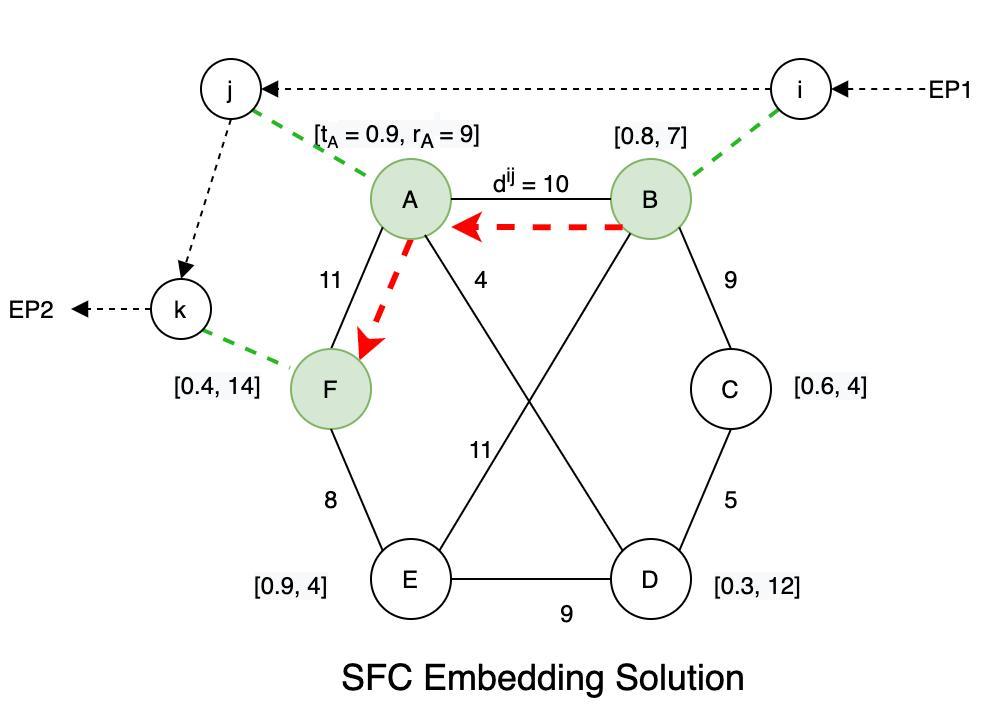}
\caption{Trust-Aware SFC Embedding Solution}
\label{fig:bwrev}
\end{minipage}
\end{center}
\end{figure*}

For instance, fig. \ref{fig: reqaug} shows an augmented graph for commodity (virtual link) $(i,j)$ in the Request Graph depicted in fig. \ref{fig:reeeeq}, that is going to be placed on the substrate network shown in fig. \ref{fig:ssssub}. Furthermore. each of the directed paths in the augmented graph depicted in fig. \ref{fig: reqaug}, that start with a \textbf{ red} edge and end with a \textbf{blue} edge, is an augmented path corresponding to commodity $(i,j)$.

\section{Problem Formulation}
\label{sec:problem}
In this section we provide the formulation for PB-SCE and PB-TASCE models as well as a k-shortest path based approximation algorithm to reduce the complexity of the optimization model while maintaining high results accuracy.  
\subsection{Path-based Model}
In order to achieve a path-based formulation which takes into account the trustworthiness of both substrate nodes and paths, we define two sets of variables to declare the formulation of the problem:
\begin{itemize}
    \item $\textbf{x}$, denotes the set of binary variables $x^i_u$ which express the assignment of VNF $i$ to substrate node $u$.
    \item $\textbf{f}$, denotes the set of continuous variables $f_p$ which express the amount of flow passing through the augmented path $p \in \mathcal{P}$ in the augmented substrate graph.
\end{itemize}
We start with a MILP formulation as follows which contains all the service requirements as hard constraints. 
\vspace{2mm}

\noindent\textbf{PB-TASCE  Objective:} 

\begin{align}
\text{\textbf{Minimize} } 
\sum_{p \in \mathcal{P}} c_p f_p + \gamma \sum_{i \in N_f}\sum_{u \in N_s} t_ux^i_u \label{obj}
\end{align}

\noindent\textbf{Placement Constraints:}

\begin{align}
& \quad \sum_{u\in N_S}x_u^i = 1, \; \quad \forall i \in N_F  \label{milp:place1}\\
& \quad \sum_{p \in \mathcal{P}^{ij}}f_p = d ^{ij}, \; \quad\forall ij \in E_f \label{place2}\\
& \quad \sum_{p \in \mathcal{P}: iu \in p}f_p \leq x^i_u M, \; \quad\forall i \in N_f, u \in N_s \label{feasflow}
\end{align}

\noindent\textbf{Trust Constraints:}

\begin{align}
& \quad(t_u-t^i)x_u^i \geq 0,  \; \quad\forall i \in N_F, \forall u \in N_S \label{milp:trust} \\
& \quad(t_p - t^{ij})f_p \geq 0, \; \quad\forall k \in E_f, p \in \mathcal{P}^{k} \label{trust2}
\end{align}

\noindent\textbf{Capacity Constraints:}

\begin{align}
& \quad\sum_{i\in N_F} g^i x_u^i \leq r_u,  \; \quad \forall u\in N_S \quad \text{ \label{milp:cap1}} \\
& \quad\sum_{p:uv \in p} f_p \leq c_{uv}, \; \quad \forall uv \in E_s\text{ \label{milp:cap2}}
\end{align}

\noindent\textbf{Domain Constraints:}

\begin{align}
& \quad x^i_u\in\{0,1\}, \; \quad \forall i\in N_F,u\in N_S  \quad \text{} \label{milp:dom2}\\
& \quad f_{p}\geq 0, \; \quad \forall p\in \mathcal{P} \text{} \label{milp:dom1} 
\end{align}


\vspace{2mm}

The objective function \eqref{obj} is the weighted sum of the flow embedding (bandwidth) and server assignment (processing) costs with $\gamma$ being the normalization factor to determine the balance between the two terms of the objective function. The processing cost corresponding to each substrate server is proportional to its trust value, i.e. the more trustworthy servers are more expensive. Constraints set \eqref{milp:place1} ensures that each request nodes is placed on one substrate node. Constraints set \eqref{place2} makes sure that the traffic demand of each request link will be allocated to this commodity using as many augmented paths as needed, while constraints set \eqref{feasflow} enforces that no flow passes through the paths that are not allowed to be used provided the node assignment policy, where $M$ is a large enough constant. 

Constraints sets \eqref{milp:trust}, and \eqref{trust2} guarantee that the trust requirements of each virtual link and each virtual node are satisfied, while constraints sets \eqref{milp:cap1}, and \eqref{milp:cap2} guarantee that the allocated CPU and bandwidth resources do not exceed the residual capacity for each substrate node and link respectively.  Constraints sets \eqref{milp:dom2}, and \eqref{milp:dom1} are the domain constraints corresponding to variable sets $\textbf{x}$ and $\textbf{f}$ respectively.
We note that removing constraints \eqref{milp:trust}, and \eqref{trust2} from the last model gives the baseline PB-SCE model. 
\subsection{Approximation Method}

We note that the PB-TASCE model cannot be used efficiently in  realistic settings with large scale networks due to not being scalable. More precisely, the complexity of the model is mostly determined by the size of path set $\mathcal{P}$, and the size of constraints set grows exponentially with the scale of the network (due to constraint \eqref{trust2}). Indeed, for a complete substrate graph, the set $\mathcal{P}$ may contain as many as  $(e|E_f|/2)(|N_s|!)$ paths \cite{path}. Even, for a sparse network graph, the size of the set of augmented paths for each virtual edge may grow exponentially in $|N_s|$. In order to tackle this issue, we modify the PB-SCE and the PB-TASCE models to contain only the $k-shortest$ augmented paths for each commodity. This will result in lower complexity at the expense of suboptimal results. Opting for different values of $k$ one can adjust the performance of the algorithm and seek for suitable value of $k$ to seek balance between complexity and result accuracy. We will explore this trade-off in detail in the evaluation section. We refer to these new models as KPB-SCE and KPB-TASCE in order.

\section{Performance Evaluation}
\label{sec:evaluation}
In this section we compare the performance of the proposed path-based models in general with the link-based model in \cite{sds19}, present the outcome of our service chain embedding scheme under both node and link trust constraints, and provide the performance evaluation results for the aforementioned approximation methods. We first provide a description of the simulation environment setup and scenarios and then proceed with presenting the evaluation results. 

\subsection{Experiment Setup}
All models and the evaluation environment are implemented in Java, including the service chain and the infrastructure topology generator. All the MILP formulations are modeled using CPLEX. For the $k-shortest$ path generator we adopted an implementation of Yen's algorithm \cite{yen}. All the experiments are conducted on an Intel Xeon processor at 3.5 GHz and 16 GB of main memory.

\noindent For the \textbf{NFV Infrastructure} we generated a 3-layer fat tree topology with 16 pods. For the evaluation setup, we used one zone of the DC with 4 pods, containing two layers of two switches and 4 servers, i.e. two servers per ToR switch, each of which having $8$ cores running at $2$ GHz. Similar to \cite{sds19}, the initial utilization and trustworthiness of each server is drawn from uniform distributions $U (0.3, 0.6)$, and $U(0.2, 1)$ in respective order. The inter-rack and ToR-to-Server link capacity are set to $16$ and $8$ Gbps accordingly. The trustworthiness of the substrate paths are randomly generated according to a uniform distribution $U(0.5, 1)$ 
 
\noindent The \textbf{SFC Requests}s were generated according to three different service chain templates as explained in great detail in \cite{sds19}. The CPU demand of each VNF is obtained from the inbound traffic rate and the VNF resource profile \cite{usenix}\cite{abujoda}. For each SFC request, the number of VNFs, and the inbound traffic demand, are generated according to uniform distributions $U(5,9)$, $U(50, 100)$. Moreover, the virtual node trust, and the virtual link trust requirement levels are both drawn from a uniform distribution $ U(0.2, 0.8)$.
 
 Similar to \cite{sds19}, for the comparison purpose, we use \textit{acceptance ratio}, \textit{CPU utilization}, \textit{bandwidth revenue\& cost}, and \textit{processing revenue\& cost} with the same definitions as provided in \cite{sds19}.

 \subsection{Evaluation Scenarios}
 We carry out two distinct sets of experiments for evaluating the performance of the proposed schemes. In the first set of experiments we compare the performance of path-based service chain embedding method to that of the link-based scheme in \cite{sds19} from different perspectives and report the results. We run the KPB-SCE model for different values of $k$ and benchmark them against the link-based method. None of the trust constraints are in place for this experiment. 
 
 The second set of experiments deal with service chain embedding under both node and link trust constraints. More precisely, this set of experiments compare the performance of the PB-TASCE model to that of the baseline PB-SCE, and PB-SCE with node constraints to highlight how the integration of trust constraints impacts the performance of the SFC embedding methods. 
 
 \subsection{Evaluation Results}
\begin{figure*}[t]
\begin{center}
\begin{minipage}[h]{0.225\textwidth}
\includegraphics[width=1\linewidth]{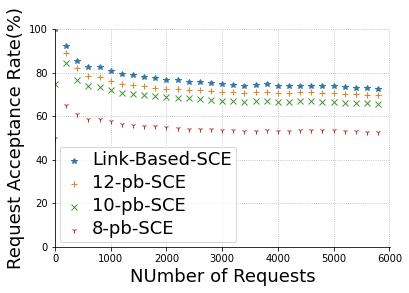}
\caption{Exp-A: Acceptance Ratio}
\label{fig:acc}
\end{minipage}
\hspace{1em}
\begin{minipage}[h]{0.225\textwidth}
\includegraphics[width=1\linewidth]{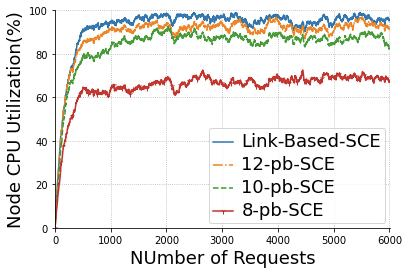}
\caption{Exp-A: CPU Utilization}
\label{fig:util}
\end{minipage}
\hspace{1em}
\begin{minipage}[h]{0.225\textwidth}
\includegraphics[width=1\linewidth]{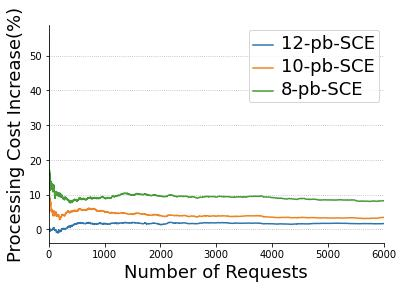}
\caption{Exp-A: Incremental CPU Revenue to Link-Based}
\label{fig:prev}
\end{minipage}
\hspace{1em}
\begin{minipage}[h]{0.225\textwidth}
\includegraphics[width=1\linewidth]{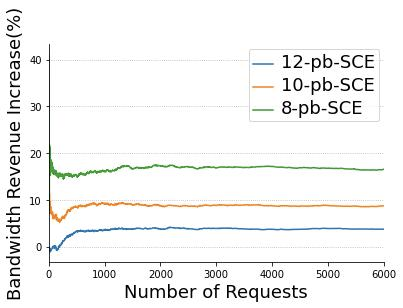}
\caption{Exp-A: Incremental BW Revenue to Link-Based}
\label{fig:bwprev}
\end{minipage}
\end{center}
\end{figure*}
\begin{figure*}[t]
\begin{center}
\begin{minipage}[h]{0.225\textwidth}
\includegraphics[width=1\linewidth]{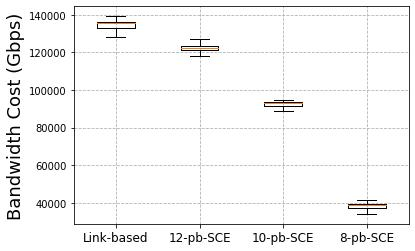}
\caption{Exp-A: Bandwidth Cost per Request}
\label{fig:bwwcost}
\end{minipage}
\hspace{1em}
\begin{minipage}[h]{0.225\textwidth}
\includegraphics[width=1\linewidth]{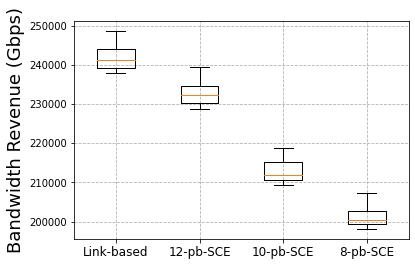}
\caption{Exp-A: Bandwidth Revenue per Request}
\label{fig:br}
\end{minipage}
\hspace{1em}
\begin{minipage}[h]{0.225\textwidth}
\includegraphics[width=1\linewidth]{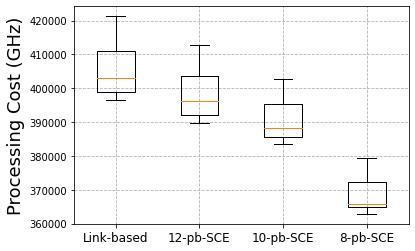}
\caption{Exp-A: Processing Cost per Request}
\label{fig:pcos}
\end{minipage}
\begin{minipage}[h]{0.225\textwidth}
\includegraphics[width=1\linewidth]{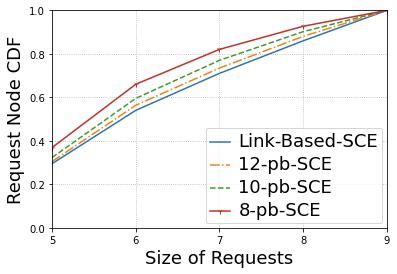}
\caption{Exp-A: CDF of Accepted Requests}
\label{fig:ccddff}
\end{minipage}
\hspace{1em}
\end{center}
\end{figure*}

 1) \textit{Experiment A:} Fig. \ref{fig:acc} shows the the comparison between the performance of link-based SCE MILP model of  \cite{sds19}, and the proposed $k$-pb-SCE algorithms, for different values of $k = 8, 10, 12$. As fig. \ref{fig:acc} depicts, as $k$ increases and more paths are included in the solution space, the performance of the $k$-pb-SCE algorithm increases. The change from $k =8$ to $k=10$ is more obvious than the change from $k=10$ to $k=12$. The $8$-pb-SCE method on average admits around $55 \%$ of the requests while $10$-pb-SCE, $12$-pb-SCE, and the link-based SCE, accept $67 \%$, $70\%$ and $74\%$ of the requests in order.
 
  Fig. \ref{fig:util} compares the CPU utilization of the substrate servers. As expected, the higher the request acceptance ratio is the higher the CPU utilization will be, as more processing resources are consumed. In steady state, in the case of the link-based SCE approach, on average more than $95 \%$ of the processing resources are consumed. The $12$-pb-SCE can almost keep up to this level, while the CPU utilization for  $8$-pb-SCE remains as low as around $70 \%$.
 
 Figures \ref{fig:prev} and \ref{fig:bwprev}, depict the percentage difference between the per-request processing and bandwidth revenue generated by the path-based approximation methods and the link-based method. By fig. \ref{fig:prev}, the processing revenue generated by the $k$-pb-SCE methods remains within $10 \%$ of the optimal link-based methods. Moreover, as fig. \ref{fig:bwprev} suggests, in steady state, the $8$-pb-SCE method provides around $14\%$ less bandwidth revenue comparing to that of the optimal link-based method. This value can be mitigated to $9\%$ and $4\%$ by taking $10$, and $12$ shortest paths for each commodity in the solution space.
 
 Figures \ref{fig:bwwcost} and \ref{fig:br} show the per-request profile of the bandwidth cost and the bandwidth revenue. Firstly, we observe a significant difference between the bandwidth cost and revenue for admitted requests which stems from the fact that different functions can be collocated on the substrate servers which will induce zero bandwidth consumption and therefore zero bandwidth cost. Moreover, we observe that the more optimal the algorithm is, the more it is successful in admitting more costly network requests. This is because when more substrate paths are injected to the solution space as the value of parameter $k$ increases, more efficient options are there for placing each request link, in the request embedding decision making process.  
 
 The box-plot for the per-request processing cost is depicted in Fig. \ref{fig:pcos}. It can be seen that the per-request processing cost for the approximation methods remain within $10\%$ of that of the link-based method confirming the observation of Fig. \ref{fig:prev}.

 Fig. \ref{fig:ccddff} shows the CDF of the number of VNFs in each SFC that is admitted by the SFC embedding mechanisms; i.e. the population of SFCs  from certain sizes that are successfully placed on the substrate network. As expected, switching from $k=8$ to $k=10$ and then $k=12$, the  profile of the admitted service chain size's converges to that of the link-based approach, which further confirms the effectiveness of our approximate embedding methods.

 \vspace{1mm}
 2) \textit{Experiment B:}  
 \begin{figure*}[t]
\begin{center}
\begin{minipage}[h]{0.225\textwidth}
\includegraphics[width=1\linewidth]{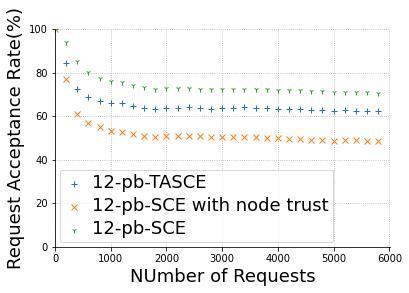}
\caption{Exp-B: Acceptance Ratio}
\label{fig:bacc}
\end{minipage}
\hspace{1em}
\begin{minipage}[h]{0.225\textwidth}
\includegraphics[width=1\linewidth]{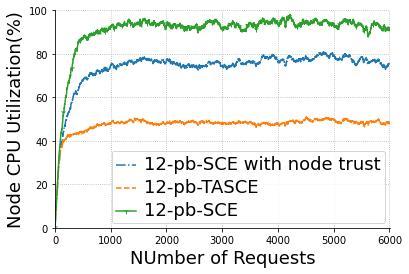}
\caption{Exp-B: CPU Utilization}
\label{fig:butil}
\end{minipage}
\hspace{1em}
\begin{minipage}[h]{0.225\textwidth}
\includegraphics[width=1\linewidth]{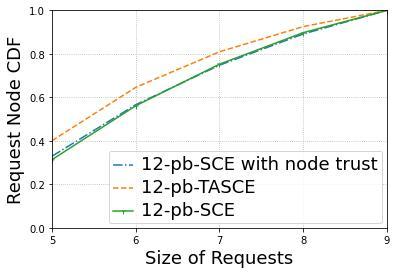}
\caption{Exp-B: CDF of Accepted Requests}
\label{fig:bcdf}
\end{minipage}
\hspace{1em}
\begin{minipage}[h]{0.225\textwidth}
\includegraphics[width=1\linewidth]{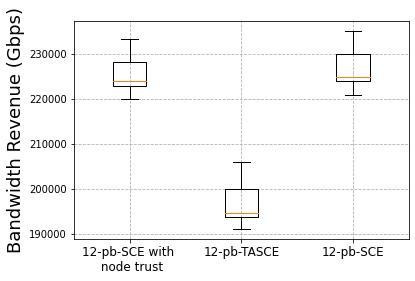}
\caption{Exp-B: Bandwidth Revenue per Request}
\label{fig:nbwprev}
\end{minipage}
\end{center}
\end{figure*}

 Fig. \ref{fig:bacc} elaborates the impact of incorporating trust into the path-based SCE model. As this figure suggests, for $k= 12$, the addition of trust requirements for the request nodes (i.e. constraint \eqref{milp:trust}) may reduce the performance of the SFC embedding method by $10 \%$ on average in the steady state. Furthermore, when the link trust requirements are integrated within the SFC embedding framework( i.e. the $12$-pb-TASCE model) the acceptance ratio diminishes by another $10\%$. 
 
 The impact of the natural decline in the acceptance ratio, when taking into account the trust constraints can be observed in the server CPU utilization profile in Fig. \ref{fig:butil} as well. One can observe that there is a $18 \%$ and a further $25 \%$ decline in the substrate CPU utilization, associated eith the addition of trust requirements for request nodes and links in the respective order. 
 We note that the performance drop caused by the node trust constraints is quite natural in that the substrate nodes with lower trustworthiness host request nodes less frequently, but the severe drop in CPU utilization due to in-existence of trustworthy substrate paths is quite more interesting; the reason being that the probability of rejecting a larger request (with more nodes and links) is higher, since due to the link trust constraints, it gets more unlikely to find feasible substrate paths for each request link when the request size increases. 
 
 To further investigate the impact of the size of requests in the embedding decision, we tested the performance of the $12$-pb-TASCE algorithm when only requests with $5$ VNFs arrive. We then repeated the same experiment for the requests of only $9$ VNFs. In the former case, we observed an increase of around $10\%$ in the CPU utilization, while in the latter this parameter dropped by around$12 \%$. 
 
 This observation is well-aligned with Fig. \ref{fig:bcdf} as well which suggests that the $12$-pb-TASCE method has a tendency to admit the smaller requests comparing to the case when there are no restrictions on the trustworthiness of the substrate paths.
 
 Finally, fig. \ref{fig:nbwprev} shows the impact of the restrictions on the trustworthiness of substrate paths, comparing to the two other cases. We observe that the per-request bandwidth revenue remains almost the same when there are only restrictions on node trustworthiness, since the set of feasible paths in the solution space does not change while when the path trust constraints are introduced the bandwidth revenue diminishes by around $15$ to $20$ percents.

\section{Related Work}
\label{sec:relatedwork}
 In this section we provide a brief review of the related works on SFC embedding and trust. 
  The literature on the SFC embedding problem is quite rich. Recently, applications of this problem have been explored in mobile edge and fog computing \cite{edge} \cite{fog} \cite{cf2}, Space-Air-Ground Integrated Networks (SAGIN) \cite{SAG}, 5G core network \cite{5gc}, multi-domain service provisioning \cite{cross1} \cite{nestor}, cloud data centers \cite{cdc}, etc. Multiple objectives and design requirements are sought when addressing the SFC embedding problem including but not limited to cost minimization \cite{cq}, resiliency \cite{raouf}, energy consumption minimization \cite{eramo}, privacy \cite{priv}, security \cite{sndr}, and trust-awareness \cite{sds19}, etc. 
 
 The trust-aware SFC embedding problem was discussed in \cite{sds19} using a link-based formulation, where only the trustworthiness of request and substrate nodes were considered in a dynamic environment. The path-based approach has already been considered in \cite{vne} in the domain of virtual network embedding where a column generation framework was proposed for the placement of virtual network functions. Among the works in the literature, our approach is more similar to \cite{vne}, and \cite{path} where the paths of substrate network are considered for the placement of network requests.
 Recently, the notion of trust has been considered in the domain of NFV and service deployment. In \cite{incorp}, the authors discuss the challenges of integrating trust within the NFV infrastructure. In the context of edge deployments and multi-domain service provisioning\cite{ced}, trust has been considered as a determining factor in deciding the most secure cloud edge deployments. The work in \cite{tez} has investigated the integration of trust into cloud by incorporating it into cloud management, different architecture components, concepts and implementation. 
\section{Conclusions}
\label{sec:conclusions}
In this paper we introduced a framework for the path-based trust-aware service chain embedding problem. We started with a baseline formulation for the path-based SFC embedding problem. Then we provided a formulation for the approximate problem by taking into account only $k$-shortest paths candidates for each virtual link. We finally incorporated the trust constrains for both virtual nodes and links and evaluated the efficiency of our algorithm through simulations and numerical results. We believe that the results accuracy and the  time complexity of the proposed path-based methods in this paper can be further improved by the development of a scheme that can dynamically add or remove the network path from the feasible solution space. A column generation framework can be used as the core of this scheme. This problem as well as providing a distributed logical framework for computing and aggregating the trustworthiness across a software-defined network, are among our future directions.

\section*{Acknowledgement}\noindent
The research was partially supported by  ONR grant N00014-17-1-262, and by grants from Lockheed Martin Corp. and Leidos Corp.



%

\end{document}